\documentclass[aps, longbibliography,reprint,prl,superscriptaddress,amsmath,amssymb,floatfix]{revtex4-2}

\usepackage[utf8]{inputenc}
\usepackage{xcolor}
\usepackage{graphicx}
\usepackage[colorlinks,allcolors=blue]{hyperref}
\usepackage{soul}

\newcommand{\BraKet}[1]{\left\langle #1 \right\rangle}
\newcommand{\MIp}{\text{MI}_{+1}}
\newcommand{\MI}{\text{MI}}
\newcommand{\MIm}{\text{MI}_{-1}}
\newcommand{\MIpm}{\text{MI}_{\pm1}}
\newcommand{\Uib}{U_{IB}}
\newcommand{\Ur}{U_{IB}/U}

\begin{document}

\title{The Bose-Hubbard polaron from weak to strong coupling}

\author{Tom Hartweg}
\affiliation{University of Strasbourg and CNRS, CESQ and ISIS (UMR 7006), aQCess, 67000 Strasbourg, France}

\author{Tanul Gupta}
\affiliation{University of Strasbourg and CNRS, CESQ and ISIS (UMR 7006), aQCess, 67000 Strasbourg, France}

\author{Guido~Pupillo}
\affiliation{University of Strasbourg and CNRS, CESQ and ISIS (UMR 7006), aQCess, 67000 Strasbourg, France}

\begin{abstract}
We investigate the zero-temperature properties of a mobile impurity immersed in a bath of bosonic particles confined to a square lattice. We analyze the regimes of attractive and repulsive coupling between the impurity and the bath particles for different strengths of boson-boson interactions in the bath, using exact large-scale quantum Monte-Carlo simulations in the grand canonical ensemble. For weak coupling, the polaron mass ratio is found to decrease around the Mott insulator (MI) to superfluid (SF) transition of the bath, as predicted by recent theory, confirming the possible use of the impurity as a probe for the transition. For strong coupling in the MI regime, instead, the impurity is found to modify the bath density by binding to an extra bath particle or a hole, depending on the sign of the polaron-bath interactions. While the binding prevent the aforementioned use of the polaron mass ratio as an MI-SF transition probe, we show that it can be used instead as a probe of the binding itself.
Our exact numerical results provide a benchmark for comparing lattice Bose polaron theories and are relevant for experiments with cold atoms trapped in optical lattices, where the presence of a confining harmonic potential can be modeled by a slowly varying local chemical potential. 
\end{abstract}

\maketitle
A mobile impurity interacting with a bath can give rise to a quasiparticle known as a polaron \cite{landau1948effective,landau1957jetp}, a paradigmatic open quantum system relevant to semiconductors \cite{franchini2021polarons},  superfluid helium \cite{bardeen1967effective}, and nuclear matter \cite{fetter2012quantum}. Recent advances in ultracold atom experiments, including quantum gas microscopy \cite{bakr2009quantum, bakr2010probing, koepsell2021microscopic, hilker2017revealing}, have enabled detailed studies of impurities in Bose-Einstein condensates \cite{grusdt2024impurities}, both in the continuum and in optical lattices described by the Bose-Hubbard (BH) model \cite{grusdt2016interferometric, camacho2019dropping, pimenov2021topological,pimenov2024polaron,koepsell2021microscopic, grusdt2018parton, wang2021higher, nielsen2021spatial, nielsen2022exact,isaule2024bound,ding2023polarons}. This has opened up the study of polaron physics in the strongly interacting regime and in the vicinity of quantum phase transitions. Lattice Bose polarons have been recently explored near the Mott insulator–superfluid (MI–SF) transition using beyond-mean-field Quantum-Gutzwiller (QGW) methods \cite{Colussi_2023}, and in combination with diagrammatic and variational quantum Monte Carlo techniques \cite{Recati_StrongInteractionPolaron_2024} at fixed particle density and in the canonical ensemble. These works demonstrate that for sufficiently weak impurity-bath interactions, polaron properties such as the energy shift and effective mass can serve as sensitive probes of the MI–SF transition and its universality class \cite{Colussi_2023}. At strong coupling, novel bound states involving the impurity and bath particles-holes excitations have been shown to emerge for fixed atoms numbers, beyond the polaron picture  \cite{Recati_StrongInteractionPolaron_2024}. A key challenge remains in providing exact results for polaron properties across all regimes of interaction, in particular for strong impurity-bath and intra-bath interactions and in the grand canonical ensemble. The latter is particularly interesting for neutral atom experiments, where the presence of  external harmonic confinement for the atoms effectively realizes a situation where local atomic densities vary across the lattice, providing effective particle reservoirs.

\begin{figure}
    \centering
    \includegraphics[width=1.0\linewidth, trim={1.5cm 6.5cm 1.5cm 6.5cm}]{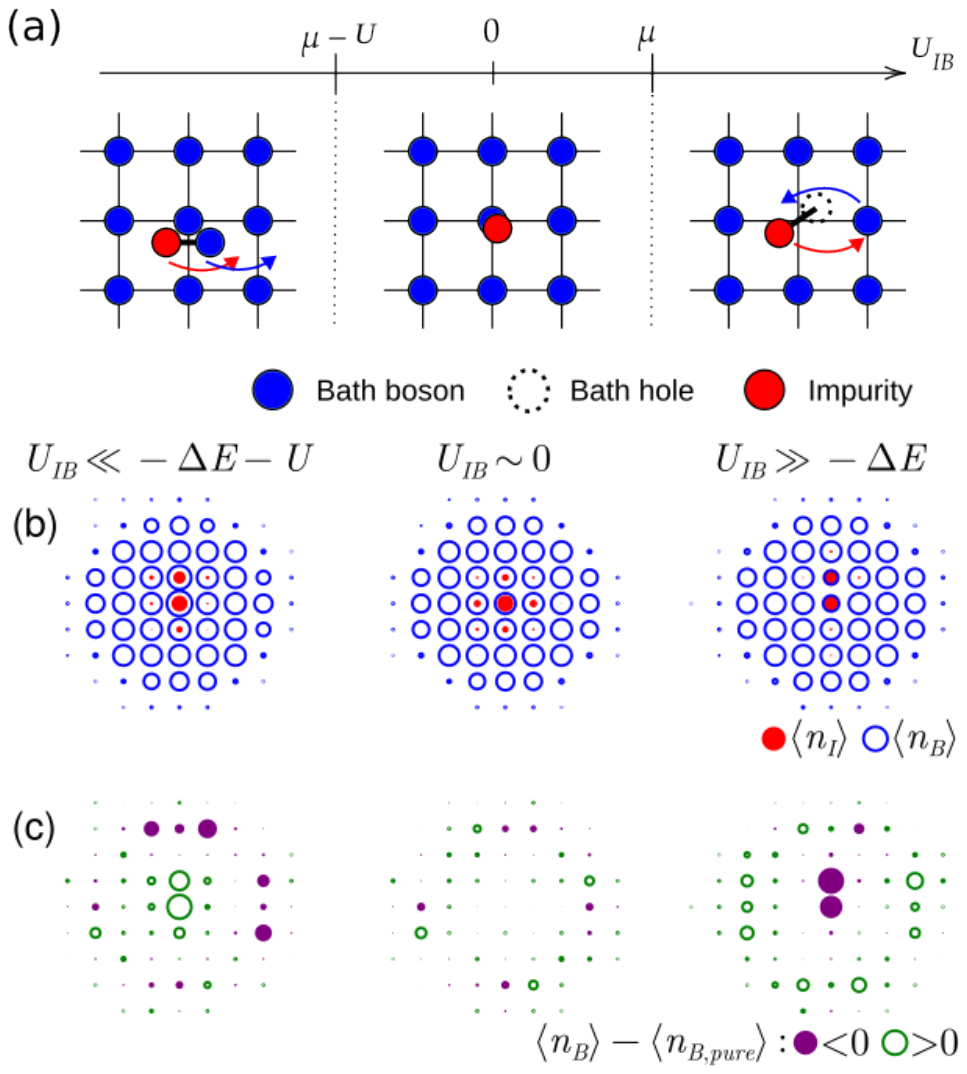}
    \caption{ (a) Schematic representation of the different states of the system in the MI regime. For $\mu-U < \Uib < \mu$, the impurity weakly perturbs the bath and remains mobile. For $\Uib < \mu - U$ ($\Uib > \mu$), the impurity binds to an extra bath particle (hole), forming a mobile bound pair that propagates via correlated (anti-correlated) hops, respectively (colored arrows). (b) QMC snapshots of a single impurity (red) in a bath of 33 bosons (blue) in a harmonic trap with $\Omega/t \approx 1.777$ at $2dt/U = 0.05$, for $\Ur = -1$ (left), 0.2 (center), and 0.5 (right). Marker areas are proportional to the local occupations $\langle n_I\rangle$ and $\langle n_B\rangle$. $\Delta E$ is the energy cost of moving a bath particle from the edge to the trap center.(c) Same parameters as in (b), but marker areas represent the deviation from the impurity-free bath, $\langle n_B\rangle - \langle n_{B,\text{pure}}\rangle$; green (purple) indicates enhanced (reduced) bath density.}
    \label{fig:snap_summary}
\end{figure}

In this work, we use exact large scale quantum Monte Carlo (QMC) simulations based on the worm algorithm \cite{prokof1998exact} to study a mobile impurity immersed in a Bose-Hubbard bath in two dimensions. Our multi-species worm algorithm (for details see Supplemental Material \cite{Note1}) can describe all regimes of impurity-bath interactions — from weak to strong, attractive to repulsive — including possible impurity-atom and impurity-hole binding.  
In order to characterize different regimes of impurity-bath interactions and possible new binding mechanisms, we focus on properties such as the polaron effective mass and the bath-induced shift as well as impurity-bath correlation functions.  For weak impurity-bath interactions, the polaron mass ratio is found to decrease around the MI-SF transition and  the polaron can be used as a probe of the transition, as predicted by \cite{Colussi_2023}. 
We quantify for what strength of interactions this picture ceases to describe the system dynamic and demonstrate that for strong interactions the impurity locally modifies the bath density resulting in a binding to an extra bath particle or a hole, depending on the sign and strength of the impurity-bath interactions (see Fig. \ref{fig:snap_summary}(a)). These binding mechanisms, which tend to suppress the polaron mass ratio and differ from the case of constant density \cite{Recati_StrongInteractionPolaron_2024}, produce distinct signatures.
We discuss the connection of these results in the grand-canonical ensemble with neutral atom experiments in the presence of an external harmonic potential. 

We consider a system comprising a single mobile impurity coupled to a BH bath on a uniform square lattice with $N=L\times L$ sites, and lattice spacing $a=1$. 
The microscopic Hamiltonian $H = H_B + H_I + H_{IB} + H_\Omega$ reads 
\begin{align}
\begin{split}
{H}_B &= -t \sum_{\langle i, j \rangle}{a}_i^\dagger {a}_{j} + \frac{U}{2} \sum_i n_{B,i}(n_{B,i} - 1),
\\ {H}_I &= -\tilde{t} \sum_{\langle i,j \rangle} {b}_{i}^\dagger {b}_{j}, \quad {H}_{IB} = \Uib \sum_i b_i^\dagger b_i a_i^{\dagger}a_i,\\ 
H_\Omega &= - \mu \sum_i n_{B,i}  + \Omega \sum_i r_i^2(n_{B,i} + n_{I,i}).\\
    \end{split}
\end{align}
Here, $H_B$, $H_I$, $H_{IB}$ and $H_\Omega$ represent the Hamiltonians for the bath, the impurity, the bath-impurity interaction, and the external harmonic trap acting on both species, respectively. The chemical potential of the bath is included in $H_\Omega$. The bosonic operators ${a}_i^\dagger$ (${a}_i$) create (annihilate) a bath particle at lattice site $i$, with ${n}_{B,i} ={a}_i^\dagger {a}_i$ the local density operator. Similarly, the impurity is described by the operators  ${b}_{i}^\dagger\;({b}_{i})$, which create (annihilate) impurity particles, with ${n}_{I,i} = {b}_{i}^\dagger {b}_{i}$. The on-site interaction energy and chemical potential of the bath are denoted by $U$ and $\mu$, respectively, while $U_{IB}$ characterizes the interaction strength between the impurity and the bath particles. The parameter $t$ ($\tilde{t}$) represents the nearest-neighbor hopping amplitude for a bath particle (impurity), $r_i$ is the distance of site $i$ from the center of the trap, and $\Omega$ represents the trap energy. We focus on impurity-bath interaction effects, assuming equal hopping, $t = \tilde{t}$. We explore scenarios involving both weak and strong impurity-bath couplings, $\Uib/t$, for both attractive ($\Uib<0$) and repulsive ($\Uib>0$) interactions. 

 Our analysis focuses on two complementary settings: (i) a homogeneous system $(\Omega = 0)$ in the grand canonical ensemble, which provides a controlled environment for estimating polaron observables and interaction-driven crossovers; and (ii) a harmonically trapped system $(\Omega \neq 0)$ at fixed total particle number, relevant to experiments, where the local chemical potential varies spatially as $\mu(r) = \mu - \Omega r^2$. For sufficiently strong interactions $U/t$, the trapped system exhibits the characteristic “wedding-cake” density profile \cite{batrouni2002mott,pupilloBraggSpectroscopyTrapped2006, mitra2008superfluid,batrouni2008canonical,rigol2009state} featuring incompressible MI cores surrounded by compressible SF shells, as shown in Fig.~1(b) and (c). 
The two settings are connected by the locally grand-canonical nature (ii), where spatial variations in the trap correspond to scanning the chemical potential in the homogeneous case. 


We begin by analyzing the polaron at $\mu = (\sqrt{2}-1)U$, which corresponds to the mean-field tip of the first Mott lobe and allows for direct comparison with recent theoretical studies \cite{Colussi_2023, Recati_StrongInteractionPolaron_2024}. For reference, the MI–SF transition in the 2D Bose–Hubbard model at unit filling along the commensurate density line has been determined via QMC to occur at $2dt/U = 0.2389(6)$ \cite{capogrosso2008monte}. In contrast to variational QMC results obtained in the canonical ensemble \cite{Recati_StrongInteractionPolaron_2024}, our numerically exact simulations in the grand-canonical ensemble \cite{Note1} show that a single impurity does not shift the transition point or affect the Mott phase properties—such as the excitation gap or compressibility—in the thermodynamic limit.


\begin{figure*}[t]
    \includegraphics[width=\linewidth]{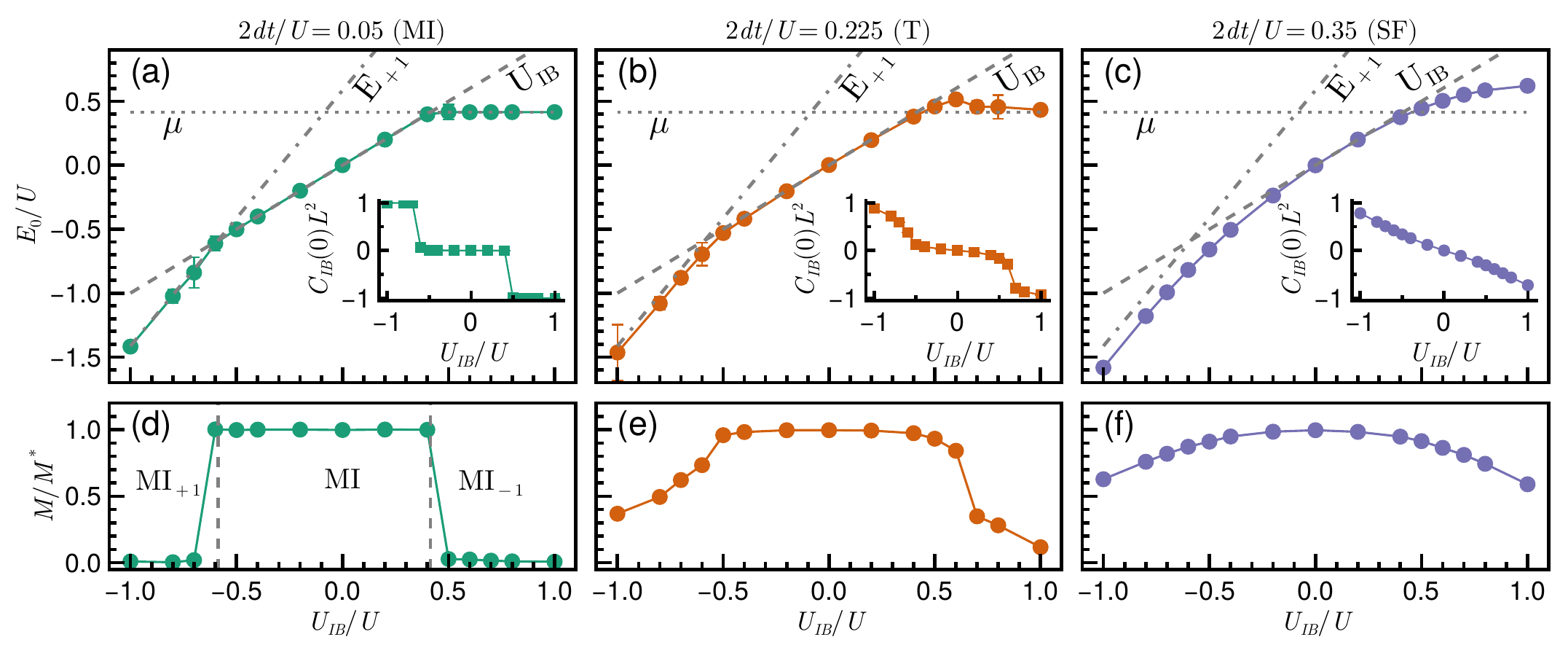}
    \caption{Top: Polaron energy shift $E_0$ as a function of impurity-bath interaction strength $\Ur$ across different bath phases: (a) Mott Insulator (MI) $(2dt/U = 0.05)$, (b) near the MI-SF transition (T) $(2dt/U = 0.225)$, and (c) superfluid (SF) $(2dt/U = 0.35)$. Solid points represent QMC estimates of $E_0$, while dashed lines show theoretical shifts. $E_{+1} = 2\Uib + U - \mu$. Insets: Impurity-bath density-density correlation $C_{IB}(r=0) $, scaled by $L^2$, showing local bath occupation changes. Bottom: effective mass ratio $M/M^*$ as a function of $\Ur$, illustrating how impurity-bath interactions modify the impurity's effective mass in different bath phases: (d) Mott Insulator (MI), (e) near the MI-SF transition (T), and (f) superfluid (SF). All plots are for an $8\times8$ lattice. Vertical dashed lines indicate $\Ur$ values at which energy level crossings occur. Error bars that are not visible are within the marker size.}
    \label{fig:bath_shift_and_mass_ratio}
\end{figure*}

We characterize the quasiparticle properties of the polaron by analyzing its energy dispersion at small momenta, $E_p(\boldsymbol{k}) = E_0 + \frac{\boldsymbol{k}^2}{2M^*} + \mathcal{O}(\boldsymbol{k}^4)$. Here, $E_0 = E_p(\boldsymbol{k} = 0)$ denotes the interaction-induced energy shift, while $M^*$ is the polaron’s effective mass, reflecting its mobility within the bath. To estimate $M^*$, we compute the impurity’s diffusion coefficient $D$ from its mean-squared displacement between imaginary times $\tau=0$ and $\tau=\beta/2$, and use the relation $M^* = 1/(2D)$ \cite{Effective_mass_1,Effective_mass_2}. Specifically,
\begin{equation}
D = \lim_{\beta \to \infty} \left\langle \left( r(0) - r\left(\frac{\beta}{2}\right) \right)^2 \right\rangle/\left(2d \cdot \frac{\beta}{2}\right),
\end{equation}
where $r(\tau)$ denotes the impurity’s position at imaginary times $\tau$, and $d = 2$ is the dimensionality of the system.
The effective mass $M^*$ is then used to estimate the bath-induced shift $E_0$ via the relation
\begin{equation}
  E_0 = E_{\text{B+I}} - E_B - E_{M^*},  \label{eq:E0}
\end{equation}
where $E_\text{B+I}$ ($E_B$) is the total energy of the system in the presence (absence) of the impurity, and $E_{M^*} = -d/M^*$ is the ground state energy of a free lattice boson of mass $M^*$. We note that  Eq. \eqref{eq:E0} includes $O(\boldsymbol{k}^4)$ contributions in $E_0$, negligible in the quasi-particle picture.
Moreover, the grand canonical ensemble allows the presence of the impurity to modify the bath's particle count with respect to a pure bath. Such a modification would be reflected by a modification of the chemical potential contribution in the total energy, and thus would also be taken into account in the value of $E_0$. We choose to use this definition as the chemical potential contribution would also appear in the case of a trapped system, even with a fixed number of particles, as discussed later.
We also compute the same-site bath-impurity density-density correlation $C_{IB}(r=0) = \BraKet{n_{B,i}n_{I,i}}_i - \BraKet{n_{B,i}}_i\BraKet{n_{I,i}}_i$, with $\BraKet{}_i$ a site and thermodynamic average. 
This observable compares the number of bath bosons at the impurity site with the site-average number of bath bosons in the system.

\emph{Polaron properties and bound states---} 
Figure \ref{fig:bath_shift_and_mass_ratio} shows QMC data for the bath-induced energy shift $E_0$ (panels (a-c)), the onsite impurity-bath correlation $C_{IB}(0)$ (insets), and the polaron mass ratio $M/M^*$ (panels (d-f)), with $M=1/(2t)$. Results are shown as a function of $\Ur$ (both attractive and repulsive) for three bath regimes: deep in the MI phase (panels (a), (d)), within the MI phase but close to the MI-SF transition (panels (b), (e)), and in the SF phase (panels (c), (f)). In particular, panel (a) shows that, for a bath in the deep MI phase, the bath-induced shift $E_0$ takes the values $E_0 \simeq E_{+1} = 2\Uib + U - \mu$ (dashed dotted line),  $\Uib$ (dashed line) and $\mu$ (dotted line) for increasing  $\Ur$ in the regions $\Uib<\mu-U$, $\mu-U <\Uib< \mu$, and $\Uib>\mu$, respectively. These regions correspond to the configurations shown in Fig. \ref{fig:snap_summary}(a): an impurity bound to an extra boson ($\MIp$, left), a free polaron ($\MI$, center), and an impurity bound to a hole ($\MIm$, right). In particular, for small values of $\Ur$ in the $\MI$ phase, $E_0 \simeq \Uib$, consistent with a mean-field picture of the impurity interacting with a homogeneous bath at $\langle n_B \rangle = 1$. For strong repulsion, $E_0 \simeq \mu$ reflects the energy cost of removing a bath particle, while for strong attraction, $E_0 \simeq 2\Uib + U - \mu$ accounts for the addition of a boson and its interaction with the impurity and the existing bath particle. The picture above is corroborated by $C_{IB}(0)$, which varies from +1 to 0 to -1 across $\MIp$, $\MI$ and $\MIm$ regions, respectively. Notably, the extra hole or particle on the impurity site reflects a true change in the bath particle number, rather than a particle–hole pair excitation \cite{Note1}. This contrasts with Ref. \cite{Recati_StrongInteractionPolaron_2024}, where the impurity cannot alter the bath occupation.

\begin{figure}[t!]
    \centering
    \includegraphics[width=1.0\linewidth]{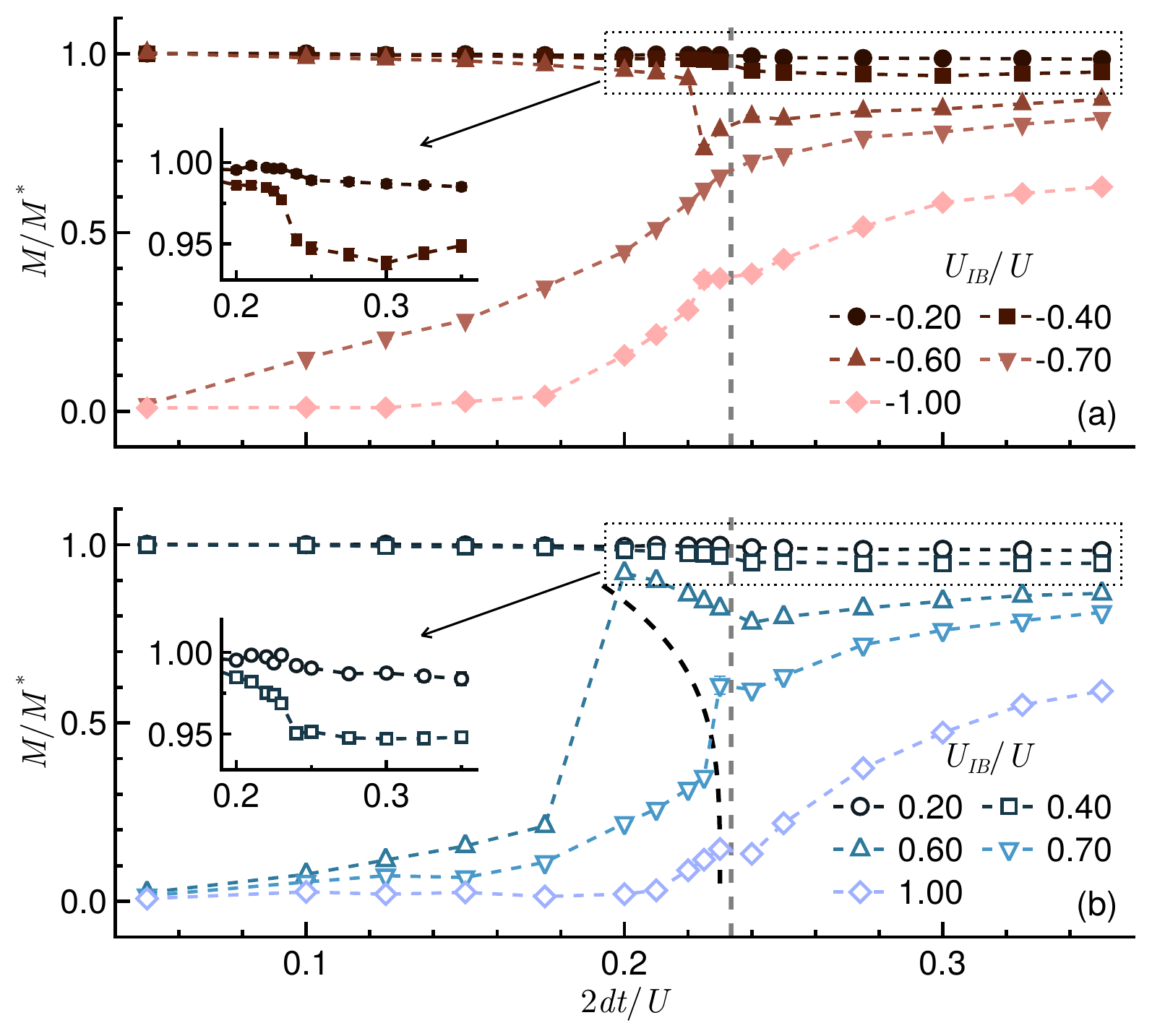}
    \caption{Polaron mass ratio $M/M^*$ as a function of  $2dt/U$ for (a) attractive and (b) repulsive branches. Different curves correspond to varying impurity-bath interaction strengths $\Ur$. Insets: zooms on black dotted boxes, showing results consistent with \cite{Colussi_2023}. Dashed grey vertical line: MI-SF transition point of the pure bath (no impurity present). Dashed black line: guide for the eye of the transition between the polaron and dimer regimes for strong repulsive interaction. All plots are for a $8\times8$ lattice. }
    \label{fig:mass_ratio_MI_to_SF}
\end{figure}
Panel (b) shows qualitatively similar features to (a), except for small deviations from the values $E_{+1}$, $\Uib$ and $\mu$ near the transition points between the different regions $\MIp$, $\MI$, and $\MIm$. The transitions are instead completely smoothed out for the case of a weakly interacting bath in the SF regime (panel (c)), where $E_0/U$ increases smoothly and monotonically from attractive to repulsive $\Ur$. The values of $E_0/U$ are qualitatively accounted for by a continuous change in  bath density on the site of the impurity and in the polaron cloud in this regime \cite{Note1}. 

The three regions are also clearly identified from the mass ratio $M/M^*$ in panel (d-f). In particular, for a bath prepared well into the MI phase (panel (d)) and for $-0.5 \lesssim \Ur \lesssim 0.5$, $M/M^*$ is close to unity, consistent with the results of \cite{Colussi_2023} for small $\Uib$.
In the $\MIpm$ regions, instead $M/M^*\simeq 0$, which is consistent with our picture above of an impurity bound to a particle and a hole, respectively. Indeed, on a 2D lattice, a free dimer consisting of two similar bosons bound by $\Uib$ is expected to have an effective hopping energy $2t^2/|\Uib|$ for $|\Uib| \gg t$ \cite{valiente2008two, mattis1986few}, resulting in an expected $M/M^*\simeq10^{-2}$, which is consistent with our QMC results. For panel (e), the transitions are again smoothed out, and the mass ratio is higher in the $\MIpm$ regions, as expected from the higher value of $2dt/U$. Panel (f) shows a continuous evolution of $M/M^*$ from strong attractive to strong repulsive interactions for a weakly interacting bath in the SF phase.
The characterization of the polaron properties across all interaction regimes and in particular the demonstration of dimer states in the $\MIpm$ regimes in the grand canonical ensemble are main results of this work.\\

\emph{MI-SF transition---} 
Figure~\ref{fig:mass_ratio_MI_to_SF} shows the polaron mass ratio $M/M^*$ versus $2dt/U$ for various values of $\Ur$. In the weak coupling limit — darker circles and squares in the attractive (panel (a)) and repulsive (panel (b)) branches — the ratio stays close to unity, consistent with the picture of a mean-field-type free polaron deep inside the MI phase. As $2dt/U$ approaches the MI-SF boundary (vertical grey dashed line), $M/M^*$ exhibits a pronounced dip (see Inset), whose depth increases with $|\Ur|$. Beyond the transition, the mass ratio rises again in the SF phase. This non-monotonic ``dip-rebound'' behavior is consistent with quantum Gutzwiller (QGW) predictions \cite{Colussi_2023}, and serves as a clear signature of the MI-SF transition in the weak coupling regime.

As $|U_{IB}/U|$ increases beyond this regime, the polaron evolves from a mobile quasiparticle into a bound dimer in the MI phase. The mass ratio $M/M^*$ then displays two distinct behaviors depending on $2dt/U$. At low hopping, the impurity binds strongly to a particle or hole excitation in the bath, forming a heavy dimer-like state. This is reflected in the strongly suppressed mass ratio observed for $|\Ur|\gtrsim 0.6$ at low $2dt/U$, signaling the $\MIpm$ region. As $2dt/U$ increases, the bound state destabilizes, and the system crosses over into the conventional MI phase, where the impurity regains mobility. This crossover, marked by curved dashed black line in the figure, occurs at progressively larger $2dt/U$ as the $|U_{IB}/U|$ increases, indicating a gradual strengthening of the bound state. This interpretation is further supported by the accompanying changes in bath density and same-site correlation $C_{IB}(0)$ \cite{Note1}. These results show that the mass ratio ceases to serve as a reliable probe of the MI–SF transition in the strong-coupling regime. However, it remains a robust indicator of impurity–bath bound-state formation.

\begin{figure}
    \centering
    \includegraphics[trim=0.5cm 0.5cm 0.3cm 1cm, clip, width=1.0\columnwidth]{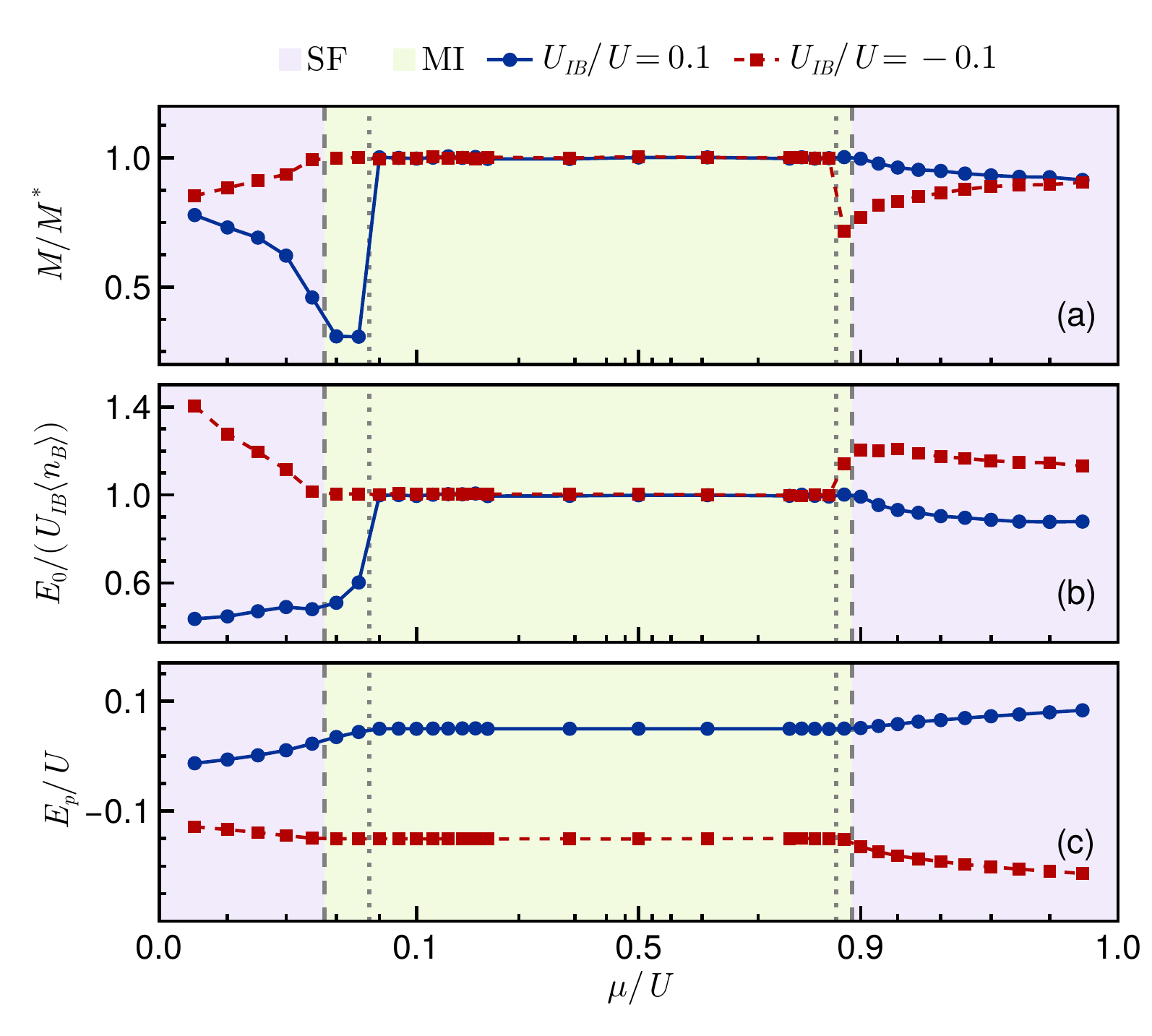}
    \caption{(a) Polaron energy $E_p/U$, (b) bath-induced shift $E_0$ normalised by the mean field shift $U_{IB}\langle n_b \rangle$, (c) mass ratio $M/M^*$ vs $\mu/U$ for $\Ur=0.1$ (solid blue line, circular markers) and $-0.1$ (dashed red line, square markers). Dashed vertical lines represent the MI-SF transition points of the bath based on estimated bath density and superfluid fraction. Dotted vertical lines are guides for the eyes for the position of the sharp features.  Here, $2dt/U=0.05$. All plots are for an $8\times8$ lattice. The $\mu/U$ scale is a pseudo-log10 centered at 0.5 and scaled by a factor 0.3.}
    \label{fig:fixed_tbyu}
\end{figure}


\emph{Harmonic trap and experimental observables---}
Having established that increasing $|U_{IB}/U|$ in a homogeneous bath drives the polaron from a mobile quasiparticle to a bound dimer, we now ask how this physics manifests in a harmonically confined system.  In the presence of a trap, spatially varying chemical potential, together with strong on-site interactions drive the bath into the familiar “wedding-cake’’ density profile \cite{batrouni2002mott,mitra2008superfluid,batrouni2008canonical,rigol2009state}. Figure \ref{fig:snap_summary}(b) shows QMC snapshots of bath and impurity densities for varying impurity–bath couplings $U_{IB}$. When the bath is initially prepared in an MI state with a thin SF shell, a sufficiently strong $U_{IB}$ locally distorts the bath density—enhancing or depleting it depending on the sign of $U_{IB}$—with compensating changes occurring at the bath boundaries (see Fig.1(c)). Although the total number of bath particles is fixed, the compressible SF shell acts as an effective reservoir, allowing for local particle-number fluctuations. In this sense the system is locally grand-canonical: moving a bath particle from radius $r_1$ to $r_2$ costs $\Delta E = -\Omega (r_1^2 - r_2^2)$, which plays the role of a local chemical-potential shift. For example, in Figure \ref{fig:snap_summary}(b), $\Delta E / U \approx -0.2$ for moving a particle from the outer edge to the center. 

Since the harmonic confinement effectively lowers the local chemical potential with increasing distance from the trap center, we simulate this radial variation by fixing $2dt/U$ and scanning $\mu/U$. Figure 4 shows QMC results for three key observables: the mass ratio $M/M^*$ (panel a), the bath-induced shift $E_0$ normalized by the mean-field value $U_{IB} \langle n_B \rangle$ (panel b), and the polaron energy $E_p = E_{B+I} - E_B$ (panel c), all as functions of $\mu/U$. These are shown for weak impurity–bath interactions $U_{IB}/U = \pm 0.08$ at fixed $2dt/U = 0.05$. In the range $0.08 \lesssim \mu/U \lesssim 0.88$, all observables remain nearly constant, corresponding to the regime where the bath is in a uniform Mott insulating state with $\langle n_B \rangle = 1$. Outside this interval, sharp features emerge. For $\mu \simeq 0.08$, the repulsive polaron (blue solid line, circular markers) exhibits a sharp cusp in $M/M^*$ and a pronounced drop in $E_0$  near $\mu/U \simeq U_{IB}/U$ in the MI phase, while the attractive polaron (red dashed line, square markers) shows smoother variations near the MI–SF boundary of the bath (indicated by vertical dashed line). Conversely, at large $\mu \simeq 0.88$, it is the attractive polaron that exhibits a sharp increase in $E_0$ and a corresponding cusp in $M/M^*$ near $\mu/U \simeq 1 - |U_{IB}|/U$, whereas the repulsive case remains smooth across the MI-SF boundary. These sharp transitions, which are not aligned with the MI–SF boundaries, are consistent with a crossover from the MI phase to the impurity-bound $\text{MI}_{\pm1}$ phases discussed earlier. The sudden suppression of $M/M^*$ signals the formation of a tightly bound bath–impurity dimer. Interestingly, the polaron energy $E_p$ (panel c) does not show corresponding discontinuities—its behavior remains smooth—implying that the abrupt changes in $M/M^*$ and $E_0$ cancel each other out in $E_p$.

In a harmonic trap, low values of the chemical potential $(\mu \lesssim 0.08)$ correspond to the outer edge of the trap and represents the superfluid (SF) shell, where the bath density  $\langle n_B \rangle < 1$. In contrast, high chemical potentials $(\mu \gtrsim 0.9)$ correspond to the SF  region between the $\langle n_B\rangle = 1$ and $\langle n_B\rangle = 2$ Mott insulating (MI) plateaus, as seen in the typical “wedding-cake” structure \cite{mitra2008superfluid, batrouni2008canonical, bakr2010probing}. These results indicate that impurity binding can occur even at weak impurity–bath coupling in a trapped system, and not just in the strong-coupling regime previously discussed in Fig.~\ref{fig:snap_summary}(b) and (c). The onset of binding is clearly reflected in the sharp changes observed in effective mass $M^*$ and bath-induced energy shift $E_0$, providing experimentally accessible signatures. In practice, by tuning the harmonic confinement—trap frequencies/curvature, depth, aspect ratio, and center offset—as well as the total atom number, one can engineer the desired bath phase at the trap center \cite{rigol2009state}; meanwhile, the impurity can be positioned independently with species-selective optical tweezers \cite{hewitt2024controlling, catani2012quantum}, enabling precise placement within targeted regions of the inhomogeneous bath.

In conclusion, we demonstrated through exact numerical analysis that increasing the impurity-bath coupling strength in a lattice Bose polaron—defined as a single impurity boson interacting with a Bose-Hubbard bath in a grand canonical ensemble—can give rise to the emergence of an MI-like phase with incommensurate density. This occurs for both attractive and repulsive interactions, as the impurity binds an additional bath particle or hole. Moreover, we demonstrated that the properties of the polaron, such as its effective mass, exhibit distinct behaviors near the MI-SF phase transition, depending on whether the impurity-bath coupling is weak or strong. We showed that the polaron mass ratio can either be used as a probe for the bath's MI-SF phase transition for weak coupling, or as a probe of the $\MIpm$ bound state for strong coupling. We argued that these results are relevant for the experimental study of the Bose lattice polaron in an harmonic trap, where the added spatial complexity make those effects possible even for a fixed number of bosons and weak or strong impurity-bath interaction. This work opens several research directions, including the study of bipolaron formation and many-polaron problems in the strong interacting regime.

\begin{acknowledgments}
{\it Acknowledgments:} We gratefully acknowledge discussions with Ragheed Alhyder and Chao Zhang. This research has received funding from the European Union’s Horizon 2020 research and innovation programme under the Marie Sklodowska-Curie project 955479 (MOQS), the Horizon Europe programme HORIZON-CL4-2021-DIGITAL-EMERGING-01-30 via the project 101070144 (EuRyQa) and from the French National Research Agency under the Investments of the Future Program projects ANR-21-ESRE-0032 (aQCess), ANR-22-CE47-0013-02 (CLIMAQS) and ANR-23-CE30-50022-02 (SIX). 
\end{acknowledgments}

\footnotetext[1]{See Supplemental Material at [URL will be inserted by publisher] for technical details on the multi-species worm algorithm used for simulation. It includes the implementation of a harmonic constraint to stabilize single-impurity simulations in the grand canonical ensemble. Additional results on the bath's compressibility, energy gap, and correlation functions are provided. Finite-size scaling analysis of winding numbers and local density shifts across the MI-SF transition are also discussed. The Supplemental Material includes Ref. ~\cite{prokof1998exact}.}

\bibliography{references}

\end{document}


\title{Supplemental Material for:\\ The Bose-Hubbard polaron from weak to strong coupling}

\author{Tom Hartweg}
\affiliation{University of Strasbourg and CNRS, CESQ and ISIS (UMR 7006), aQCess, 67000 Strasbourg, France}

\author{Tanul Gupta}
\affiliation{University of Strasbourg and CNRS, CESQ and ISIS (UMR 7006), aQCess, 67000 Strasbourg, France}

\author{Guido~Pupillo}
\affiliation{University of Strasbourg and CNRS, CESQ and ISIS (UMR 7006), aQCess, 67000 Strasbourg, France}

\maketitle

\subsection{Details on the adaptation of the worm algorithm}

We study the lattice Bose polaron using a multi-species adaptation of the worm algorithm \cite{prokof1998exact}, where each atomic species is treated on a separate worldline, and the bath-impurity interaction is taken into account in the move probabilities of the respective worms. Allowing the opening of a worm for each species simultaneously is required to ensure fast convergence, especially in the presence of possible impurity-bath bound states where correlated or anti-correlated bath-impurity hopping are present, which results in strong multi-modal distribution for a single worm. To ensure detailed balance, both worms are opened and closed at the same time, and each move consist in choosing randomly a specie and moving the associated worm.
Furthermore, the worm algorithm naturally works in the grand canonical ensemble. To treat the single impurity, we do not take into account worldlines with an inadequate number of impurities. To ensure that this scheme works efficiently, we add a fictitious term $\tilde{H}=\gamma(N_I-N_T)^2 - \mu_IN_I$ in the Hamiltonian $H$, where $N_I=\sum_i n_{I,i}$ is the total number operator of impurities, $N_T(=1)$ is the targeted $N_I$, $\gamma$ is a real positive parameter, and $\mu_I$ is a chemical potential. As the only closed worldlines taken into account verify $N_I-N_T=0$, $\tilde{H}$ only consequence is to shift the targeted system's energy by $-\mu_I$, which can be removed manually afterward. The harmonic term $\gamma(N_I-N_T)^2$ helps reducing the need to fine tune the chemical potential of the impurity. The open worldlines can be reweighted to compensate for $\tilde{H}$ and be used to study the Green functions exactly. Finite size calculations are performed with inverse temperature $\beta=L\times L$, where $L$ is the lattice size.

\subsection{Insulating properties of the bath.}
\begin{figure}[t!]
    \centering
    \includegraphics[width=1.0\linewidth]{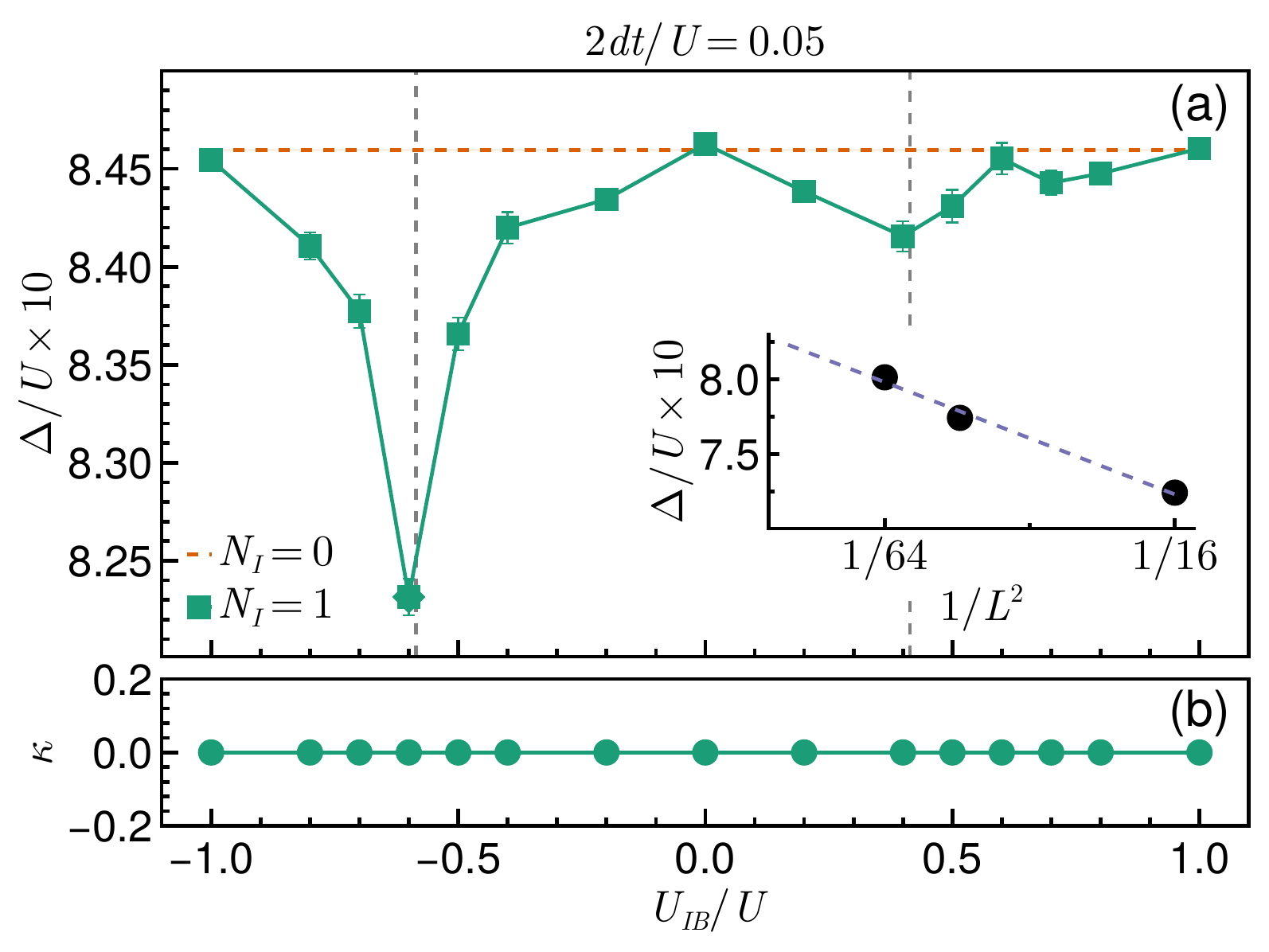}
    \caption{(a) Thermodynamic limit of the Energy gap $\Delta$ in function of interaction strength $U_{IB}/U$, for a system with 1 and 0 impurity ($N_I$), estimated by extrapolating from system sizes of $4\times4$, $6\times6$ and $8\times8$. The color band around the $N_I=0$ data represents estimated uncertainty. Vertical dashed lines are situated at $U_{IB}/U=1-\mu/U$ and $\mu$. Inset: example of thermodynamic scaling with the data from the point $U_{IB}/U=-0.6$ indicated with a star marker (b) Compressibility $\kappa$ of the bath in function of $U_{IB}/U$ for a system size of $8\times 8$. Error bars not visible are within the marker size.}
    \label{fig:energy_gap}
\end{figure}

\begin{figure}[h]
    \centering
    \includegraphics[width=\linewidth]{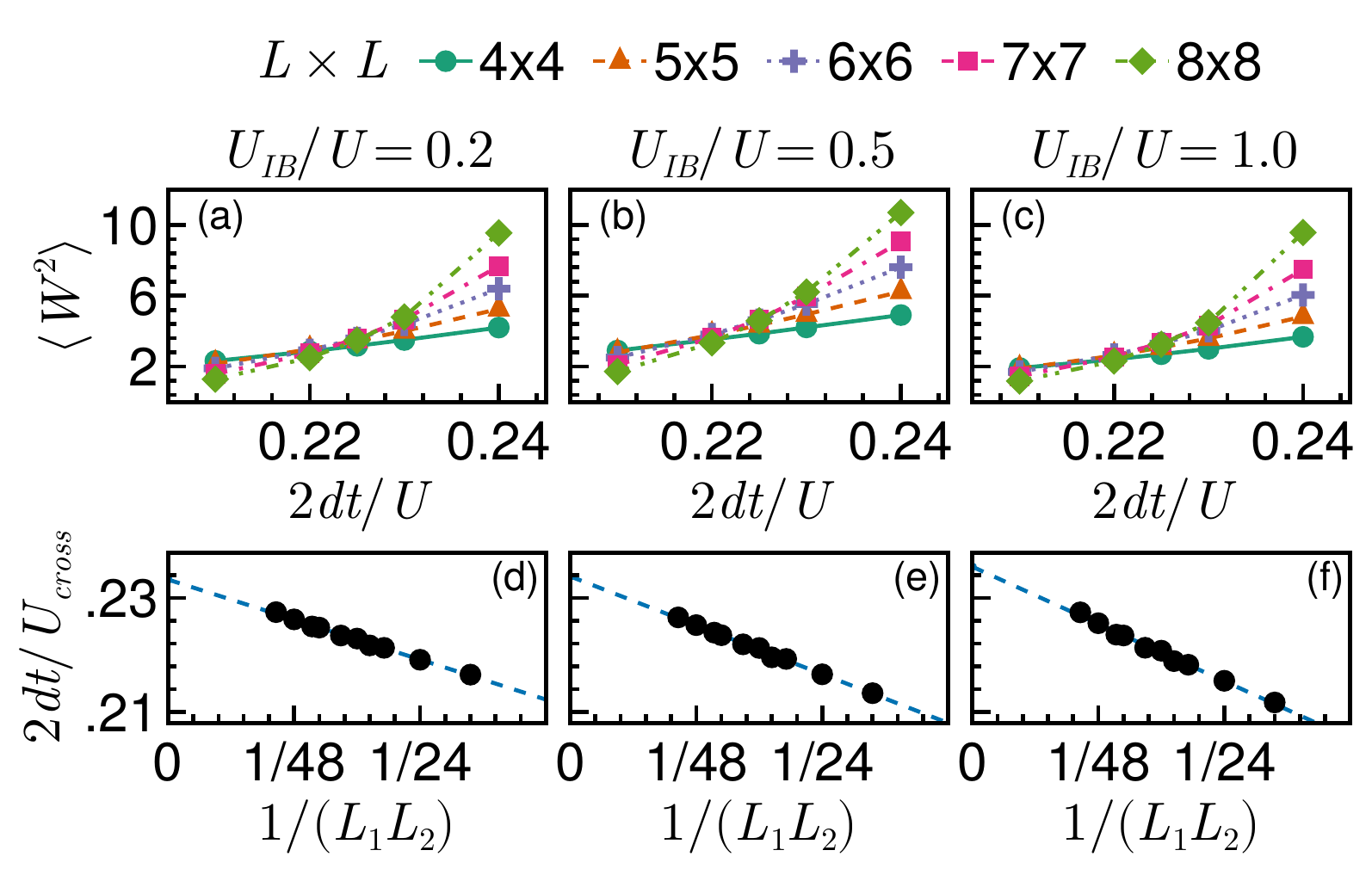}
    \caption{(a-c) Bath windings squared in function of $2dt/U$, for $U_{IB}/U = 0.2$ (a), $0.5$ (b) and $1.0$ (c) for different system sizes ($L$ = 4,5,6,7,8). (d-f) scaling of the windings crossing point for systems with respective sizes $L_1$ and $L_2$, for $U_{IB}/U = 0.2$ (d), $0.5$ (e) and $1.0$ (f).}
    \label{fig:critical_point}
\end{figure}

As the overall density of the bath in the $\MI$ phase is not commensurate in the strong impurity-bath coupling regime, we further explored its properties by computing the compressibility $\kappa$ and the energy gap $\Delta$ of the system, presented in Figure \ref{fig:energy_gap}. The compressibility was retrieved by analyzing the Monte-Carlo statistics of the density as $\kappa = \beta/L^2\left(\BraKet{N_B^2} - \BraKet{N_B}^2\right)$ where $N_B$ is the total number of bosons in the bath. The energy gap was estimated by fitting the exponential decay in complex time $\tau$ of the zero-momentum Green function $\mathcal{G}(k=0,\tau)$. At the value of $2dt/U=0.05$, the compressibility is shown to be null while the energy gap is non-zero for all values of $U_{IB}$ in the range $-1$ to $1$, indicating that the bath is still in an MI-like phase, despite the addition of a particle or a hole in the $\MIp$ and $\MIm$ regimes.

Moreover, we computed the MI-SF transition point in the homogeneous system at fixed $\mu = (\sqrt{2}-1)\times U$, for $U_{IB}/U=0.2$, $0.5$, and $1.0$, using an analysis of the crossing of the spatial windings of the bath, for different system size. Th data and analysis are presented in figure \ref{fig:critical_point}. The values obtained are respectively $2dt/U=.23(3)$, $.23(3)$ and $.23(5)$, showing no significant effect of $U_{IB}/U$ on the MI-SF transition.

\subsection{Impurity back-action on the bath and finite size effect.}
\begin{figure}
    \centering
    \includegraphics[width=1.0\linewidth]{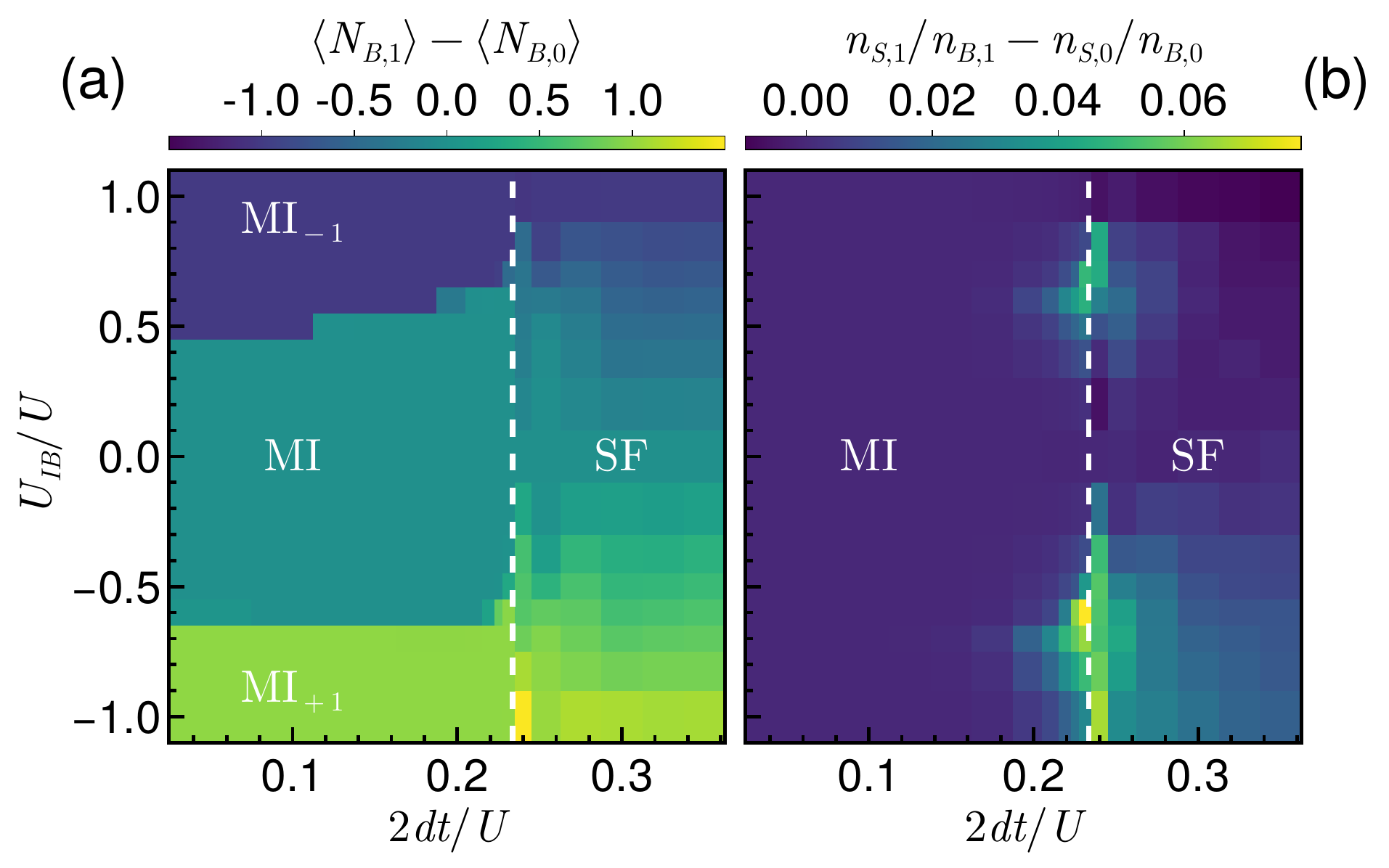}
    \caption{Color maps illustrating the effects of impurity-bath interaction strength $ U_{IB}/U$ and hopping parameter $2dt/U$ on the bath properties. (a) The difference in the number of bath bosons $\langle N_{B,1} \rangle - \langle N_{B,0} \rangle$ between systems with one and zero impurities highlighting impurity-induced local density variations across the Mott insulator (MI) and superfluid (SF) phases.  (b) Superfluid fraction difference $n_{S,1}/n_{B,1} - n_{s,0}/n_{b,0}$, indicating how the impurity affects superfluidity. Both color maps are for a system size of $8\times8$ and present an estimated error on the order of (a) $10^{-2}$ and  (b) $10^{-3}$. The dashed vertical line marks the MI-SF phase transition of the pure bath.}
    \label{fig:density_and_spf_modif}
\end{figure}

\begin{figure}
    \centering
    \includegraphics[width=1.0\linewidth]{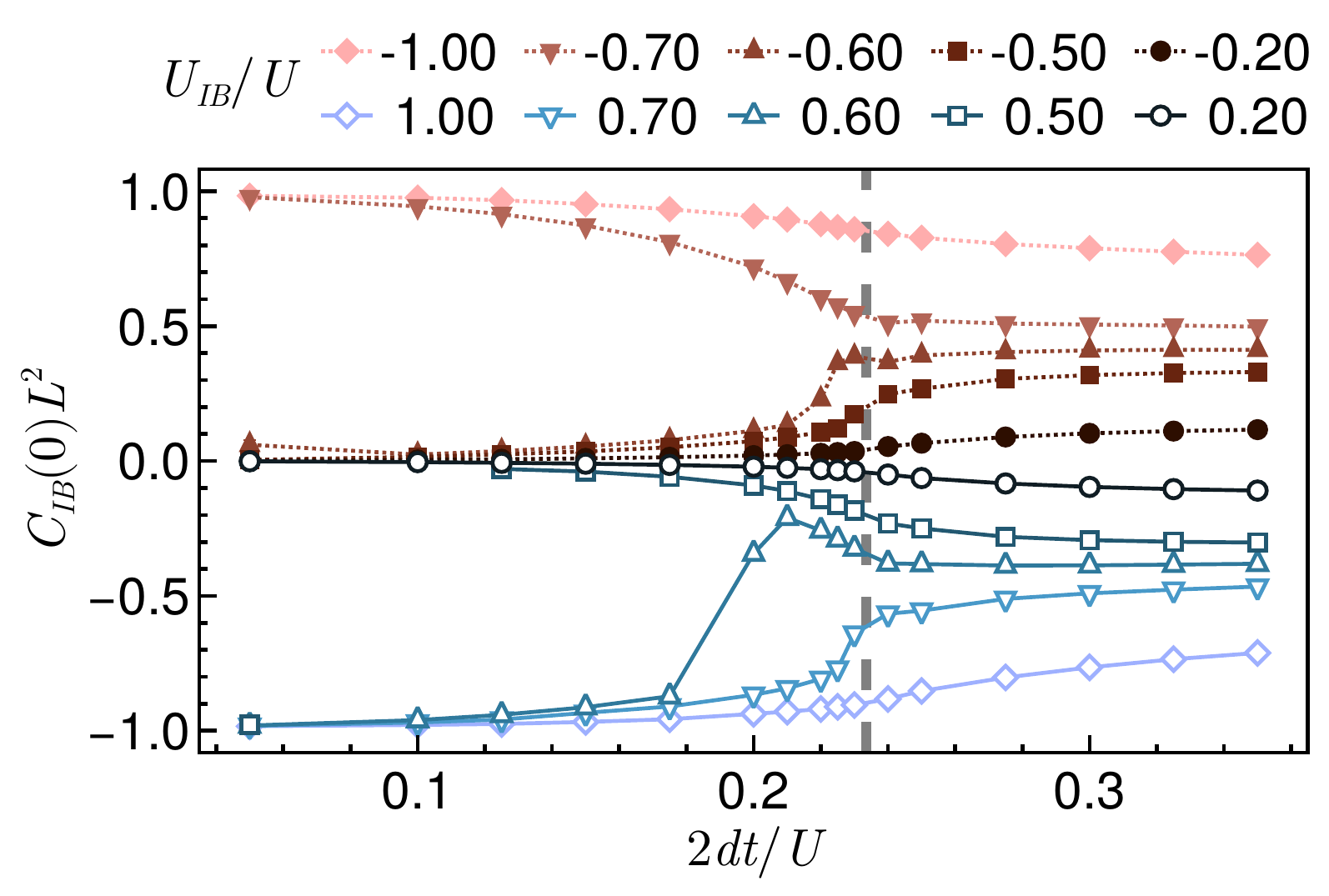}
    \caption{On-site density-density correlation $C_{IB}(r = 0)$ across the MI to SF transition of the bath, for different impurity-bath interaction strength $\Ur$. The dashed vertical line marks the MI-SF transition point of the pure bath.}
    \label{fig:density-density_correlation_onsite}
\end{figure}

As discussed in the main text, the impurity can have noticeable back-action on the homogeneous bath in the strong bath-impurity interaction. Figure \ref{fig:density_and_spf_modif}(a) shows the QMC data difference in number of bath boson in the system with an impurity and from a pure system, in function of $\Ur$ and $2dt/U$. In the MI phase ($2dt/U\lesssim0.23$), we clearly distinguish between three regions in function of $\Ur$. Indeed for $\Ur\gtrsim\mu/U$, we see that the addition of the impurity remove one bath particle. For $\Ur\lesssim\mu/U-1$, the presence of the impurity add an extra bath particle. In between, the impurity do not affect the particle count. Those three regions match with the described $\MIm$, $\MI$ and $\MIp$ phases, and show that the particle or hole bound to the impurity is indeed added by the presence of the impurity, and not taken from another site. We also see that the boundaries between the $\MI$ and $\MIpm$ are not totally horizontal and changes with the value of $2dt/U$, as discussed in the main text. For the SF phase ($2dt/U\gtrsim0.23$), the dependence on $\Ur$ is smoother. We notice that the strongly repulsive impurity ($\Ur=1.0$) still removes around 1 bath particle, while the strongly attracting impurity ($\Ur=-1.0$) adds more than 1 bath particle ($1.17(4)$), as stated in the main text.

Figure \ref{fig:density_and_spf_modif}(b) shows the difference in superfluid fraction between a system with one impurity and a pure system. We see, close to the MI-SF transition, a slight increase in the superfluity of the bath around $|U_{IB}/U| \approx 0.6$. However, the appearance of superfluidity before the thermodynamic limit of the transition point is due to finite-size effects, as shown by the non modified MI-SF transition in the thermodynamic limit.

Figure \ref{fig:density-density_correlation_onsite} shows the data for the correlation function $C_{IB}(r=0)$ as a function of $2dt/U$ for attractive (solid markers) and repulsive (hollow markers) $\Ur$. For weak $\Ur$ (darker colors), the values are close to zero in the MI, and become finite and scaling with $\Ur$ in the SF. However, for strong $\Ur$ (lighter colors), the values are close to $1$ or $-1$ in the MI, as expected from the $\MIp$ and $\MIm$ phases, but get closer to zero in the SF. We point out that $C_{IB}(r=0)$ only quantifies local bath density change relative to a pure bath. To have the total number of bath particle on the impurity site, we have to take into account the varying density of the pure bath in the SF phase. Indeed, as we keep $\mu/U$ fixed, the bath density increases linearly from 1 at the MI-SF transition to almost $1.1$ at $2dt/U=0.35$. Thus, deep in the SF ($2dt/U=0.35$), the number of bath particle on the strongly attractive impurity ($\Ur=-1.0$, solid diamond) site is kept around 1.

As stated in the main text, we notice that this back-action can qualitatively explain part of the deviation of the bath induced shift $E_0$ in the SF phase presented in \ref{fig:bath_shift_and_mass_ratio}(c).
For $\Ur \simeq 1$, for example, $E_0$ approaches the value $E_0/U \simeq \mu/U+0.2(0)$, while for values $\Ur \simeq -1$, $E_0/U$ approaches $E_{+1}/U-0.(4)*\mu/U$. The bath density on the sites of the strongly repulsive ($\Ur=1.0$) impurity is about $0.37$, compared to $0$ in the $\MIm$ phase, resulting in an expected increase in the bath shift of $0.37U$. The $\mu$ contribution is present as the number of extra holes in the bath is still around $1$, as shown in figure \ref{fig:density_and_spf_modif}. For the strongly attractive case ($\Ur=-1.0$), the situation is reversed. In this regime the bath density on the impurity site is still around $1$, while the polaron cloud gets an increase in the number of extra bath bosons of $0.17(4)$ with respect to the MI case. Those extra bath bosons are expected to contribute $-0.17\mu$ to $E_0$. These values agree qualitatively with the QMC results.

\bibliography{references}
\bibliographystyle{apsrev4-2}